\documentclass[prl, aps, twocolumn, superscriptaddress, 10pt]{revtex4-1}

\usepackage[utf8]{inputenc} 
\usepackage[T1]{fontenc} 
\usepackage[english]{babel} 
\usepackage{lmodern} 
\usepackage{geometry} 
\usepackage{setspace} 
\usepackage{indentfirst} 
\usepackage{graphicx} 
\usepackage{hyperref} 
\hypersetup{
	colorlinks=true,
	linkcolor=blue,
	citecolor=blue,
	urlcolor=blue
}
\usepackage{natbib} 
\usepackage{verbatim} 
\usepackage[pangram]{blindtext} 

\usepackage{amsmath, amsthm, amscd, amssymb} 
\usepackage[all,cmtip]{xy} 
\usepackage{listings} 
\usepackage{physics} 
\usepackage{lipsum}

\def \dif {\mathrm{d}}
\def \cm {\mathrm{c}}
\def \sm {\mathrm{s}}

\makeindex

\begin{document}
\title{Overdamped dynamics of a Brownian particle levitated in a Paul trap}

\author{Gerard P. Conangla}\email{gerard.planes@icfo.eu}
\affiliation{ICFO Institut de Ciències Fotòniques, Mediterranean Technology Park, 08860 Castelldefels (Barcelona), Spain}

\author{Dwight Nwaigwe}
\email{dwightnwaigwe@math.arizona.edu}
\affiliation{Program in Applied Mathematics, University of Arizona, Tucson, AZ 85721-0089, USA}

\author{Jan Wehr}
\email{wehr@math.arizona.edu}
\affiliation{Department of Mathematics, University of Arizona, Tucson, AZ 85721-0089, USA}

\author{Raúl A. Rica}\email{rul@ugr.es}
\affiliation{ICFO Institut de Ciències Fotòniques, Mediterranean Technology Park, 08860 Castelldefels (Barcelona), Spain}
\affiliation{Universidad de Granada, Departamento de Física Aplicada, Facultad de Ciencias, 18071 Granada, Spain}
\date{\today}

\begin{abstract}
We study the dynamics of the center of mass of a Brownian particle levitated in a Paul trap. We mostly focus on the overdamped regime in the context of levitodynamics, comparing theory with numerical simulations and experimental data from our Paul trap. We provide an exact analytical solution to the stochastic equation of motion, expressions for the standard deviation of the motion, and thermalization times by using the WKB method under two different limits. Finally, we prove the power spectral density of the motion can be approximated by that of an Ornstein-Uhlenbeck process and use the found expression to calibrate the motion of a trapped particle.
\end{abstract}

\maketitle

\section{Introduction} 
One of the strengths of cavity optomechanics, and in particular of the optomechanical Hamiltonian, is the large range of different experimental systems to which it can be applied~\citep{Aspelmeyer2014}. The role of the mechanical oscillator can for instance be taken by a membrane resonator: these microfabricated systems have allowed ground-state cooling of a mechanical mode in a macroscopic system for the first time~\citep{Oconnell2010,Teufel2011,Chan2011}. Nonetheless, the formalism is the same for other systems of mechanical oscillators interacting with light. Levitated systems are no different~\citep{chang2010cavity,romero2010toward,millen2019optomechanics,delic2019motional}.

Research in levitodynamics has concentrated most of its efforts on the study of the COM motion of levitated objects. For this reason, optical tweezers and silica particles have become the most common trapping duple in the field~\citep{Li2010,Gieseler2012}, due to its simple experimental requirements. However, the possibility to levitate micro- and nano-particles with optically active internal degrees of freedom, investigate materials hosting emitters, or study richer Hamiltonians demands the use of alternative trapping mechanisms. Quadrupole ion traps offer a different solution that, unlike optical fields, does not heat trapped particles through optical absorption---a major cause of particle degradation in vacuum---while providing large trapping volumes and deep potential wells~\citep{Kuhlicke2014}. Several groups have demonstrated their utility by studying different particle materials, including plasmonic nanoparticles~\citep{Schell2017flying} and diamonds with embedded NV centers~\citep{Kuhlicke2014, Delord2017thermometry, Delord2017, conangla2018motion}. They have also been used in cavity cooling experiments~\citep{millen2015cavity, fonseca2016nonlinear}, and are candidates for hybrid systems of nanoparticles coupled to ions~\citep{bykov2019direct} and all-electrical systems~\citep{goldwater2019levitated}.

The invention of the quadrupole ion trap in the late 50s had a tremendous impact in physics~\citep{Paul1990}, since it allowed the trapping of ions and, thus, gave rise to the manipulation of individual quantum objects under well-controlled conditions for the first time~\citep{wineland2013nobel}. Naturally, after the first demonstrations of ion trapping appeared, the study of the ion dynamics in quadrupole traps increased at a vertiginous pace~\citep{blatt1986brownian, joos1989langevin, maitra2019far}. Despite this, due to the very restrictive experimental conditions required for ion trapping, the pressure regimes dominated by damping and Brownian noise~\citep{izmailov1995microparticle, holtkemeier2016dynamics} did not receive much attention until optical tweezers detection techniques were adapted to nanoparticles levitated in Paul traps~\citep{Nagornykh2015,alda2016trapping}. Furthermore, many recent studies with micro and nanoparticles in Paul traps were mainly focused on the determination of the trapped object's optical properties~\citep{krieger2012exploring,davies2012time,bell2014single,howder2015thermally} or in detecting small signals from internal degrees of freedom~\citep{Neukirch2013,Kuhlicke2014,Delord2017,conangla2018motion}. These experiments would benefit from a good understanding of the particle dynamics, since minimizing the variance of the position greatly improves the particles' signal collection.

To bridge this gap, we present a theoretical and experimental study of the dynamics of a particle levitated in a Paul trap starting from its stochastic differential equation. We provide the exact solution to the equation, expressions for the standard deviation $\sigma_y$ of the motion and thermalization times by using the WKB method~\citep{bender85m} under two different limits. We show that a naïve description based on the overdamped approximation---typically used by the optical tweezers community~\citep{jones2015optical}---is not valid in the case of a Paul trap. Finally, we apply the found expressions to prove the motion power spectral density (PSD) can be approximated by that of an Ornstein-Uhlenbeck (OU) process~\citep{oksendal2013stochastic}. We complete our study by a thorough comparison of all the theory results with simulations and provide an example of an experimental application by calibrating the motion of a trapped particle.

\section{Solution of the equation of motion}

The COM motion along an axis of a classical particle levitated in a quadrupole trap can be described~\citep{Major2006} by the stochastic differential equation of motion
\begin{align}\label{eq:motion}
m \ddot{y} + \gamma \dot{y} - \epsilon \cos \omega_\text{d} t\cdot y = \sigma \eta(t),
\end{align} 
where we have neglected DC electric fields. Here, $m$ is the particle mass~\citep{ricci2019accurate}, $\gamma$ is the constant damping due to the interaction with residual gas molecules, $\omega_\text{d}$ is the trap driving frequency, epsilon is defined as
\begin{align*}
\epsilon \triangleq\frac{QV}{d^2},
\end{align*}
where $Q$ is the particle charge, $V$ the trap voltage and $d$ is the characteristic size of the trap (related to the distance between electrodes), and finally $\sigma \eta(t)$ is a stochastic force, with $\eta(t)$ being a unit intensity Gaussian white noise and $\sigma = \sqrt{2 k_\text{B} T \gamma}$~\citep{Kubo1966}. We notice that at the pressure regimes of this work (ambient to low vacuum), the quadratic region of the electric potential in the trap is much larger than the particle confinement. Therefore, the dynamics along the different axes are uncoupled, and the 1D equation that we have just presented is a good description for the motion along all of them~\citep{gieseler2013thermal}.

An explicit expression for $y(t)$ can be obtained as follows. We introduce the normalized damping $\Gamma = \gamma/m$ and perform the change of variables $y = e^{-\frac{\Gamma}{2}t}y_1$, obtaining:
\begin{align}\label{eq:paultrap}
m \ddot{y}_1 - \left(\frac{m\Gamma^2}{4} + \epsilon \cos \omega_\text{d} t\right)y_1 = \sigma e^{\frac{\Gamma}{2}t}\eta(t).
\end{align}
Introducing a dimensionless time $\tau=\omega_\text{d}t/2$ and defining 
\begin{align*}
y_2 = \frac{m\omega_\text{d}}{2}y_1,
\end{align*}
we get to:
\begin{align}
\begin{split}
\ddot{y}_2 - \left(\frac{\Gamma^2}{\omega_\text{d}^2} + 2\cdot \frac{2\epsilon}{m\omega_\text{d}^2} \cos (2\tau)\right)y_2 = \\
 \frac{2\sigma}{\omega_\text{d}} e^{\frac{\Gamma}{\omega_\text{d}}\tau}\eta\left(\frac{2\tau}{\omega_\text{d}}\right).
\end{split}
\end{align}

Notice that with $a = - \Gamma^2/\omega_\text{d}^2, \quad q = 2\epsilon/m\omega_\text{d}^2$ we obtain the Mathieu equation at the left hand side of the equation, plus a noise term:
\begin{align}
\begin{split}
\ddot{y}_2 + \left(a - 2\cdot q \cos (2\tau)\right)y_2 = \\
 \frac{2\sigma}{\omega_\text{d}} e^{\frac{\Gamma}{\omega_\text{d}}\tau}\eta\left(\frac{2\tau}{\omega_\text{d}}\right).
\end{split}
\end{align}

We can now rewrite this equation as a first order linear system with $Y = (y_2, v_2)^\top$, where $v_2 = \dot{y_2}$. In Itô's notation~\footnote{In the noise differential we used $ \eta\left(\frac{2\tau}{\omega_\text{d}}\right) \frac{2}{\omega_\text{d}}\dif \tau = \eta(t)\dif t = \dif W_t = \sqrt{\frac{2}{\omega_\text{d}}} \dif W_\tau$. The time change formula for Itô integrals was used, see Oksendal, Theorem 8.5.7}.
\begin{align}\label{eq:complete}
\begin{split}
\dif Y = 
\begin{pmatrix}
0 & 1 \\ 
-a + 2q \cos (2\tau) & 0
\end{pmatrix}
\cdot Y \dif \tau + \\
\begin{pmatrix}
0\\  
\sqrt{\frac{2}{\omega_\text{d}}}\sigma e^{\frac{\Gamma}{\omega_\text{d}}\tau}
\end{pmatrix}
\cdot \dif W_{\tau}.
\end{split}
\end{align}
Here, $W_\tau$ is a Wiener process---the time integral of a white noise $\eta(t)$ driving the equation of motion
\begin{align}
W_t = \int_0^t\eta_s\,\dif s.
\end{align}
The fundamental matrix solution of the associated homogeneous noise-free system is given by the Mathieu cosine and sine functions~\footnote{For a detailed discussion of the Mathieu equation and functions' properties, see \url{http://dlmf.nist.gov/28.2}.} (see \hyperref[sec:mathieu]{Appendices}):
\begin{align*}
\Phi(\tau) = 
\begin{pmatrix}
\cm_{a,q}(\tau) & \sm_{a,q}(\tau) \\ 
\dot{\cm}_{a,q}(\tau) & \dot{\sm}_{a,q}(\tau)
\end{pmatrix}.
\end{align*}

With Liouville's formula~\footnote{It is also called the Wronskian of the system} we see that the determinant of this matrix is 1. Therefore its inverse is
\begin{align}
\Phi^{-1}(\tau) & = \det \Phi(\tau)^{-1}\cdot
\begin{pmatrix}
\dot{\sm}_{a,q}(\tau) & -\sm_{a,q}(\tau) \\ 
-\dot{\cm}_{a,q}(\tau) & \cm_{a,q}(\tau) 
\end{pmatrix} \nonumber\\
& =
\begin{pmatrix}
\dot{\sm}_{a,q}(\tau) & -\sm_{a,q}(\tau) \\ 
-\dot{\cm}_{a,q}(\tau) & \cm_{a,q}(\tau) 
\end{pmatrix}
\end{align}
and hence we can solve the complete system (see \hyperref[sec:explicit_solution]{Appendices}). Solving for the particle's position we obtain
\begin{widetext}
\begin{align}
y_2\left(\frac{2\tau}{\omega_\text{d}}\right) = 
\begin{pmatrix}
\cm_{a,q}(\tau) & \sm_{a,q}(\tau)
\end{pmatrix}
\cdot
\begin{pmatrix}
y_2(0)\\ 
v_2(0)
\end{pmatrix}
+
\begin{pmatrix}
\cm_{a,q}(\tau) & \sm_{a,q}(\tau)
\end{pmatrix}
\cdot
\int_0^{\tau}
\sqrt{\frac{2}{\omega_\text{d}}}\sigma e^{\frac{\Gamma}{\omega_\text{d}}u}\cdot
\begin{pmatrix}
-\sm_{a,q}(u)\\ 
\cm_{a,q}(u)
\end{pmatrix}
\dif W_u
\end{align}
which is the solution of~\eqref{eq:complete}. Undoing the time change and setting $y_2 = \frac{\omega_\text{d} m}{2}e^{\frac{\Gamma t}{2}}y$ we obtain
\begin{align}\label{eq:fullsolution}
\begin{split}
y(t) =  
e^{-\frac{\Gamma t}{2}} \begin{pmatrix}
\cm_{a,q}(\omega_\text{d} t/2) & \sm_{a,q}(\omega_\text{d} t/2)
\end{pmatrix}
\cdot
\begin{pmatrix}
y(0)\\ 
\frac{2}{\omega_\text{d}}v(0) + \frac{\Gamma}{\omega_\text{d}}y(0)
\end{pmatrix}
+ \\
\frac{2}{\omega_\text{d} m}
\begin{pmatrix}
\cm_{a,q}(\omega_\text{d} t/2) & \sm_{a,q}(\omega_\text{d} t/2)
\end{pmatrix}
\cdot
\int_0^{t}
\sigma e^{\frac{\Gamma}{2}(r - t)}\cdot
\begin{pmatrix}
-\sm_{a,q}(\omega_\text{d} r/2)\\ 
\cm_{a,q}(\omega_\text{d} r/2)
\end{pmatrix}
\dif W_r
\end{split}
\end{align}
\end{widetext}
which is an explicit analytical solution of equation~\eqref{eq:motion} and coincides with the solution provided in Ref.~\cite{park2012thermal}. Note that the initial condition terms decay exponentially and do not affect the long term behavior of the equation. 

\section{Approximate solution and PSD}

Even though the solution in~\eqref{eq:fullsolution} is exact, it is difficult to obtain useful parameters from it---such as its moments' asymptotic behavior---without numerical methods. We thus provide an alternative approximate solution by following the WKB procedure to find the short and long term behavior of the physically relevant process variance $\langle y^2(t) \rangle$. 

With the WKB computations, one seeks a solution of the form
\begin{align*}
y(\tau) = e^{-\tau/2\kappa}\exp(\frac{S_0}{\kappa} + S_1 + \ldots)
\end{align*}
as a series in terms of $\kappa$, known as the ``small parameter''. Since the choice of $\kappa$ has a certain degree of arbitrariness, we provide calculations for two $\kappa$ alternatives. We then follow our analysis by further approximation, dropping small terms of the series, and find that both choices lead to a result of the same form. Numerical simulations serve as a final check for the validity of the expressions.

For the first series, we use the particle mass $m$ as the small parameter $\kappa$. In the second case, we rewrite eq.~\eqref{eq:motion} as
\begin{align}\label{eq:newequation}
\kappa^2 \ddot{y} +\kappa \dot{y} - \beta' \cos(2 \tau) y=  \frac{m \sigma}{\gamma^2} \sqrt{\frac{\omega_\text{d}}{2}}  \eta(\tau),
\end{align}
which is a reformulation of~\eqref{eq:motion} where $\tau = \frac{\omega_\text{d} t}{2}$ as before, $\beta'= \frac{m \epsilon}{\gamma^2}$ and $\kappa = \frac{m \omega_\text{d}}{2 \gamma}$ is the new small parameter.  

At this point we introduce a set of parameter values, in accordance with our experimental setup (see \hyperref[sec:parameters]{Appendices}), for which $\kappa \ll 1$, and thus ensure the validity of the method. The derivations provided are completely general except when an approximation relies on the specific values of the chosen parameters, in which case it is explicitly indicated. 

Stating the problem with the initial conditions $y(0) = v(0) = 0$ for greater clarity, the solution of the equation of motion is then found to be approximately (see \hyperref[sec:wkb]{Appendices}):
\begin{align}\label{eq:wkb_solution}
y(t) \approx \frac{\sigma }{ \gamma}\int_0^te^{\lambda\left(t-s\right)}\,\dif W_s.
\end{align}
As before, $W_t$ is a Wiener process and $\lambda$ is the larger of the two characteristic Floquet exponents. $\lambda$ can be thought of as a thermalization rate: the time it takes for the particle to reach ``equilibrium'' is proportional to $1/|\lambda|$. However, one should recall that the Paul trap equation is non-autonomous and, therefore, the concepts from the theory of stationary processes, when they apply, are only approximations.

The value of $\lambda$ depends on the way the limit is taken and how many terms of the series expansion are considered (see \hyperref[sec:wkb]{Appendices}), but in both cases it is a negative number with a small absolute value that can be approximated as
\begin{align}\label{eq:lambda}
\lambda \approx - \frac{m\epsilon^2 }{ 2\gamma^3}.
\end{align}
Our numerical tests show that this is a good approximation when the equation parameters are similar to actual nanoparticle experiments~\citep{Nagornykh2015,millen2015cavity,alda2016trapping,Delord2017,conangla2018motion}, for pressures between a few milibars and ambient.

According to the It\^o isometry~\citep{oksendal2013stochastic}, the variance of the approximate solution equals 
\begin{align}
\mathbb{E}[y^2(t)] = \left(\frac{\sigma }{ \gamma}\right)^2\int_0^te^{2\lambda\left(t-s\right)}\,\dif s = \left(\frac{\sigma }{ \gamma}\right)^2\frac{1 - e^{2\lambda t}  }{ 2\left|\lambda\right|}.
\end{align}

\begin{figure*}[t]
\includegraphics[width=0.8\textwidth]{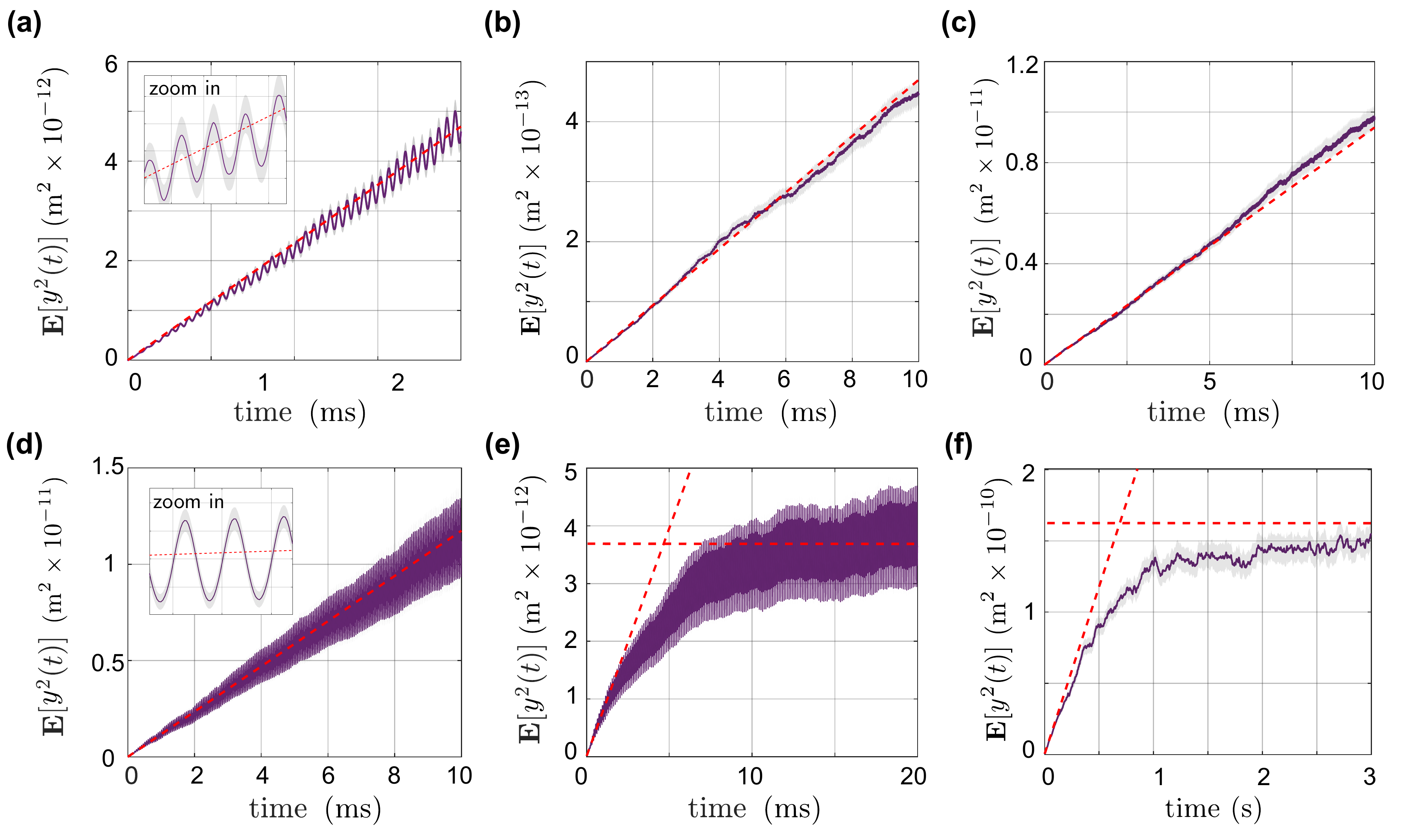}
\caption{\textbf{Variance and reheating} Comparison of numerical simulations of the variance (in purple) with the analytical expressions (in red dashed lines) found in the theory section. Sub-figures (a) to (d) show simulations of $\mathbb{E}[y^2(t)]$ (starting from $(x,v) = (0,0)$ at $t=0$) for $t \ll 1/|\lambda|$ in a range of typical experimental values, proving the validity of the expressions for common nanoparticle experiments. The insets in (a) and (d) show the oscillatory behaviour of the variance due to the trap driving, along with confidence intervals of $\pm 1$ standard error (these margins can't be seen in some of the figures, due to the oscillations being much larger than the uncertainty). Figures (b) and (c), although similar at first sight, show very different speeds in a reheating experiment. The last two sub-figures, (e) and (f), show simulations of $\mathbb{E}[y^2(t)]$ for $t > 1/|\lambda|$: thermalization around an equilibrium value can be observed. The two figures portray a case where the oscillations dominate (e) and a case where the oscillations are small with respect to the equilibrium value (f).  Again, the gray area around the variance (in purple) represents $\pm 1$ standard error. \textbf{Simulation details}: $T = 295$ K, Paul trap with $f = 20$ kHz, $V=1000$ V. (a) Pressure $p =101$ mbar, particle with diameter $d=200$ nm, $Q=100\text{ e}^+$. (b) Pressure $p =1010$ mbar, particle with $d=1$ $\mu$m, $Q=500\text{ e}^+$. (c) As (a) but $p=1010$ mbar. (d) As (c) but $d = 40$ nm. (e) $d = 200$ nm, $Q=500\text{ e}^+$, $p = 300$ mbar and $V = 2000$ V. (f) As (e) but $p = 1010$ mbar and $V = 1000$ V.}
\label{fig:2}
\end{figure*}

It follows that
\begin{align}\label{eq:long}
\mathbb{E}[y^2(t)] \simeq \frac{\sigma^2 }{ 2\gamma^2 \left|\lambda\right|}
\end{align}
for $t \gg 1/|\lambda|$ and that the variance converges to this value in the limit $t \to \infty$. For short times the diffusion term dominates and it grows linearly
\begin{align}\label{eq:short}
\mathbb{E}\left[y(t)^2\right] \approx \left(\frac{\sigma }{ \gamma}\right)^2t.
\end{align}
The actual duration of this regime depends on the exact physical parameters. By taking values from common nanoparticle experiments~\citep{Nagornykh2015,millen2015cavity,alda2016trapping,Delord2017,conangla2018motion}, we find it should lie in the $10^{-2}$ s to several seconds range (see Fig.~\ref{fig:2}).

Expression~\eqref{eq:wkb_solution} is approximately equal to the stationary solution of the OU equation~\citep{oksendal2013stochastic}
\begin{align}
\dif x_t = \lambda x_t + \frac{\sigma }{ \gamma}\,\dif W_t,
\end{align}
which also describes optically trapped particles in the overdamped regime. Therefore, the covariance of the process $y(t)$ is approximately equal to the covariance of the stationary OU process, that is
\begin{align}
\mathbb{E}\left[y(t)y(u)\right] \approx \frac{\sigma^2}{2\gamma^2 \left|\lambda\right|}e^{\lambda\left|u-t\right|}.
\end{align}
By the Wiener-Kinchine theorem, the spectral density of $y(t)$ is
\begin{align}\label{eq:psd}
S_y(\omega)& = \int_{-\infty}^{\infty}e^{-i\omega t}\frac{\sigma^2 }{ 2\gamma^2 \left|\lambda\right|}e^{\lambda|t|}\,\dif t \\
& = \left(\frac{\sigma }{ \gamma}\right)^2\frac{1 }{ \lambda^2 + \omega^2} = \frac{2k_\textrm{B}T/\gamma}{ \lambda^2 + \omega^2} 
\end{align}
We provide a more detailed argument by using explicit bounds on the error of the WKB approximation (see \hyperref[sec:psd]{Appendices}). From~\eqref{eq:psd}, we clearly see that our parameter $|\lambda|$ corresponds to the cutoff frequency of the trap~\cite{roldan2014measuring}.

\section{Experimental setup}

\begin{figure}
\begin{center}
\includegraphics[width=0.48\textwidth]{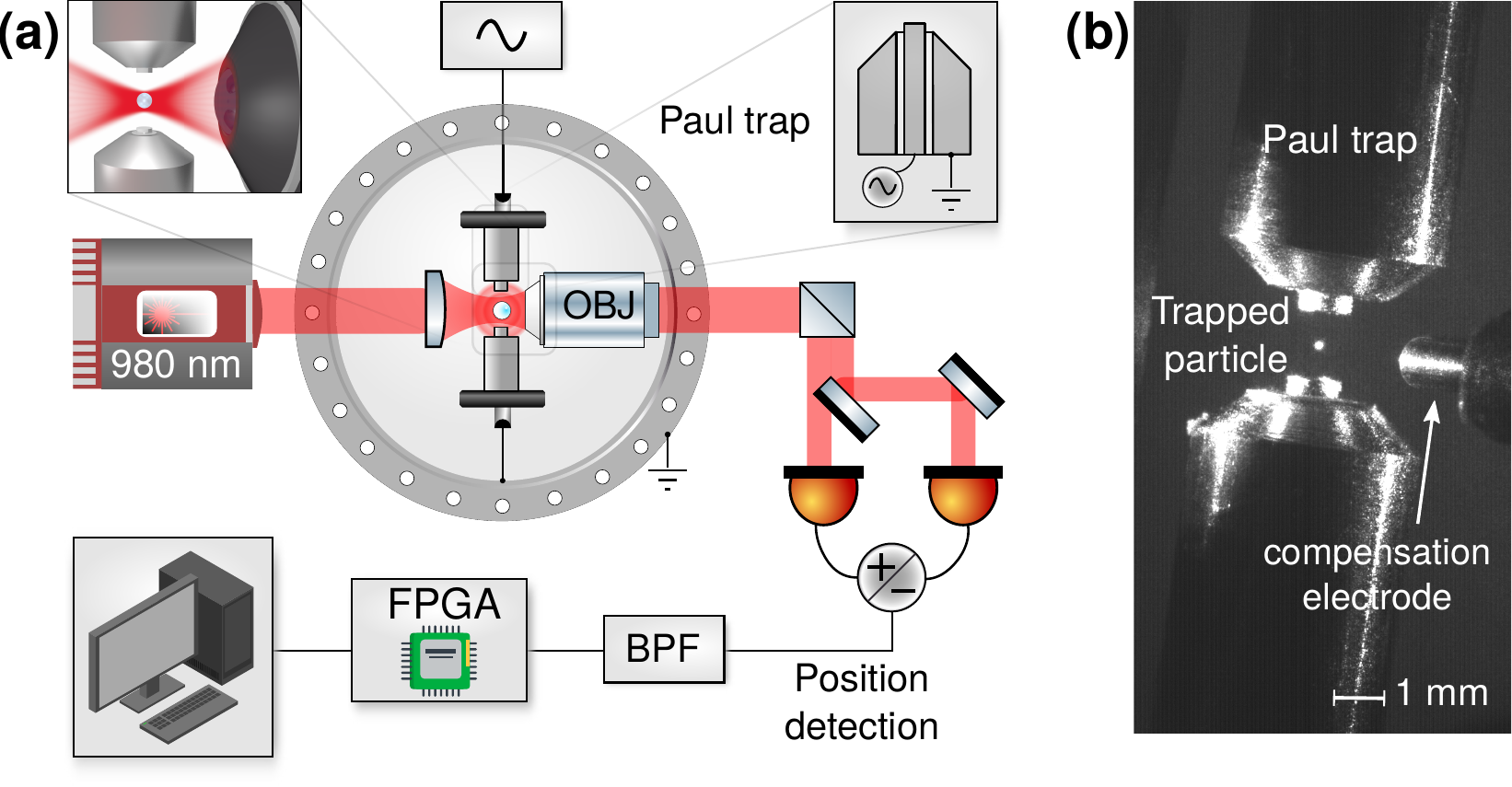}
\caption{\textbf{Experimental setup} (a) Setup sketch. A Paul trap, driven by a high voltage sinusoidal signal, levitates a charged nanoparticle that is illuminated from the left with a 980 nm diode laser. The scattered light is collected with an objective (OBJ) and sent to a quadrant photodiode to detect the motion. The measured signals are band-pass filtered (BPF) and sent to an FPGA, where they are further pre-processed and sent to a computer. (b) Picture of the trap inside the vacuum chamber viewed from above, showing the end-cap Paul trap electrodes and one of the compensation electrodes.}
\label{fig:1}
\end{center}
\end{figure}

The experimental setup is displayed in Fig.~\ref{fig:1}. We use a rotationally symmetric end-cap Paul trap to levitate charged nanoparticles (silica and polystyrene). It is designed to provide optical access and a linear electric field in a large volume around the trapping region. The Paul trap is made of two assembled steel electrodes separated by 1.4 mm mounted on a ceramic holder. The oscillating field is created by a high voltage signal (sinusoidal with frequency between $1-30$ kHz, amplitude of $0.6-2$ $\text{kV}_\text{pp}$).

The nanoparticles, originally suspended in ethanol, are loaded into the trap with a custom made electrospray at ambient pressure. This ensures that particles are highly charged ($50 < n < 1000$ of net e$^+$ charges in this study, depending on the particle). The Paul trap is mounted on a piezo stage inside a vacuum chamber, giving access to pressures below ambient conditions. A laser, focused with an aspheric lens with low effective numerical apperture ($\text{NA} < 0.1$), illuminates the particle. Its scattering signal is collected in forward detection with an objective (0.8 NA) and directed to a quadrant photodiode, giving signals proportional to the particle motion in the 3 perpendicular directions $x(t)$ (parallel to trap axis), $y(t)$ (gravity direction) and $z(t)$ (beam propagation). The signals are sent to an FPGA and recorded in a computer.

\section{Results and discussion} 

We have simulated~\footnote{The code and libraries can be freely downloaded from \href{https://github.com/gerardpc/sde\textunderscore simulator}{https://github.com/gerardpc/sde\_simulator}.} the stochastic differential equation to check the validity of the approximate solution~\eqref{eq:wkb_solution} and the expression~\eqref{eq:psd} of the motion's PSD. We have also measured position traces from a real trapped particle, computed the PSD, and compared the results to the model to calibrate the position signal (i.e., find the factor to convert a signal in volts to meters). Finally, we have studied the dependence of $\sigma_y \triangleq \sqrt{\mathbb{E}[y^2(t)]}$ on the experimental parameters $\omega_\text{d}$ and $Q\times V$ for long $t$ to ascertain the best regimes for trapping in terms of particle confinement.

The behavior of $\mathbb{E}[y^2(t)]$ for common nanoparticle experimental parameters is displayed in Fig.~\ref{fig:2}. Subfigures (a) to (d) compare the diffusive regime ($t \ll 1/|\lambda|$) of the process variance to the expression~\eqref{eq:short} for different parameter choices. In all of our numerical experiments $\mathbb{E}[y^2(t)]$ closely follows the $\frac{\sigma^2}{\gamma^2}t$ trend. However, in some of them a small oscillation at $\omega_\text{d}$ can be seen, which is not present in our model due to the approximations of the WKB method. These realizations of the process start with initial conditions in the origin $x(0) = 0$, $v(0) = 0$, and represent the behavior we would expect from a cooled nanoparticle (for instance with feedback cooling through the compensation electrodes~\citep{conangla2019optimal}) after the cooling is turned off at $t = 0$. 

Fig.~\ref{fig:2} (e) and (f) show the long term trend of the position variance ($t > |\lambda|$). The diffusive regime and the asymptotic ``equilibrium'' value found in~\eqref{eq:long} are marked with red dashed lines, and the simulations are found to follow these closely (the higher the number of averaged traces, the smaller the variance of the simulation and the better the resemblance with the model). We found a surprisingly long thermalization time for the particle motion to reach equilibrium, that we have verified with numerical simulations. Depending on the experimental conditions, it might take seconds (as is the case of Fig.~\ref{fig:2} (f)) or even tens of seconds for common values extracted from the literature. This is in stark contrast with optical dipole traps. At ambient or low vacuum pressure regimes, a cooled optically trapped nanoparticle reaches equilibrium in the timescale of microseconds. The results indicate that one should be cautious when taking data from a trapped particle on a Paul trap, since the process memory might be longer than expected and the results could still be affected by the initial conditions.

Another interesting finding is that, in this regime, $\mathbb{E}[y^2(t)]$ does not converge to the value expected from the equipartition theorem if one naïvely uses the effective potential approximation for quadrupole traps. For a simple harmonic oscillator, $\mathbb{E}[y^2(t)] = k_\text{B}T/m\Omega^2$, and using the expression for the secular frequency of a weakly damped Paul trap~\citep{whitten2004high}
\begin{align}\label{eq:secular}
\Omega = \frac{\omega_\text{d}}{2}\sqrt{\frac{q^2}{2}-\frac{\Gamma^2}{\omega_\text{d}^2}}
\end{align}
one finds that
\begin{align}
\mathbb{E}[y^2(t)] = \frac{k_\text{B}T }{m\Omega^2} = \frac{8k_\text{B}T}{m\omega_{\text{d}}^2\left(q^2-\frac{2\Gamma^2}{\omega_{\text{d}}^2}\right)}.
\end{align}
From Eq.~\eqref{eq:secular}, we see that the secular approximation breaks down if $2\Gamma^2/\omega_\text{d}^2> q^2$, as it is typically the case at ambient pressure. However, Eq.~\eqref{eq:long} shows---with the support of numerical simulations---that
\begin{align}\label{eq:equilibrium}
\mathbb{E}[y^2(t)] \simeq \frac{\sigma^2 \gamma}{m\epsilon^2}=\frac{8k_\text{B}T}{mq^2\omega_{\text{d}}^2}\left(\frac{\Gamma^2}{\omega_\text{d}^2}\right).
\end{align}
Notice that the right-hand expression does not really depend on $\omega_\text{d}$, since $q \propto 1/\omega_\text{d}^2$. A refined approximation to the variance was given in Ref.~\citep{izmailov1995microparticle}:
\begin{align}\label{eq:izmailov}
\mathbb{E}[y^2(t)] \simeq \frac{8k_\text{B}T}{mq^2\omega_{\text{d}}^2}\left(1+\frac{\Gamma^2}{\omega_\text{d}^2}\right)I_0^2\left[\frac{q}{\sqrt{1+\frac{\Gamma^2}{\omega_\text{d}^2}}}\right],
\end{align}
where $I_0$ is the zeroth order modified Bessel function.

\begin{figure*}[t]
\includegraphics[width=0.77\textwidth]{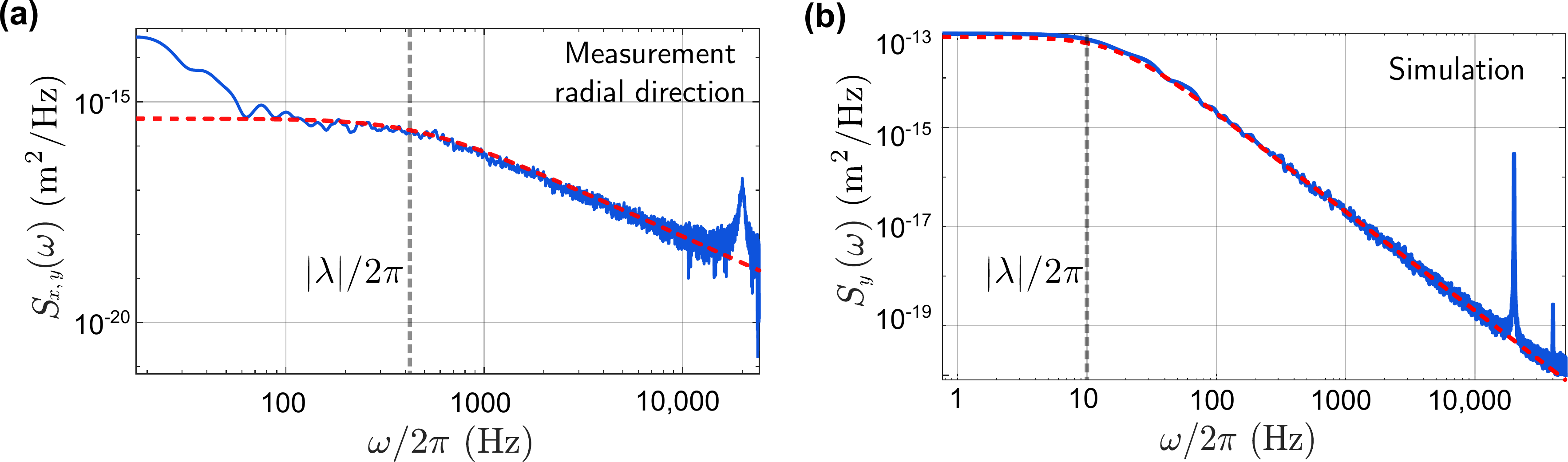}
\caption{\textbf{Motion spectral densities} Comparison of measurements and numerical simulations of the power spectral densities (PSD), in blue, with the analytical expressions (in red dashed lines) found in the theory section. 
(a) Measurement of the PSD $S_{x,y}(\omega)$ in the trap radial directions $x, y$ for a polystyrene particle of 243 nm in diameter (Paul trap with $V = 1000$ V and $f = 20$ kHz, $p = 100$ mbar $\pm 30\%$ gauge uncertainty). A fit of the expression found in~\eqref{eq:psd} is superimposed, with amplitude and corner frequency as free parameters. The fitted corner frequency, $|\lambda| / 2\pi \simeq 420\text{ Hz}$, is marked with a grey dashed line in the figure. From eq.~\eqref{eq:lambda} we recover a value of $\epsilon$ compatible with the pressure, voltage and frequency of the trap. 
(b) Simulation of the process with the expression in~\eqref{eq:psd} superimposed (dashed red line) but no free parameters (using the values of the simulation: $V = 1000$ V, $f = 20$ kHz, $p = 1010$ mbar, $d = 200$ nm, resulting in $|\lambda|/2\pi \simeq 10$ Hz). Excellent agreement between model and simulation is found except at $\omega = \omega_\text{d}$, since the trap driving is suppressed in the Ornstein-Uhlenbeck model.}
\label{fig:3}
\end{figure*}

The approximation in~\eqref{eq:equilibrium}, of course, is not valid for any choice of values in parameter space, since we have used several approximations based on our own experiment. Nonetheless, it is a good approximation for nanoparticles in Paul traps at ambient and similar pressure regimes, as long as the condition $\Gamma\gg\omega_\text{d}$ is fulfilled and the stability criteria of the Mathieu equation are met~\cite{izmailov1995microparticle, nasse2001influence}, i.e., $q$ is not too large (in particular, $q\lesssim\sqrt{1+\Gamma^2/\omega_\text{d}^2}$, since in this case $I_0\simeq 1$). Eq.~\ref{eq:izmailov} has a minimum at $q_\text{min}=1.518\sqrt{1+\Gamma^2/\omega_\text{d}^2}$, which finds the optimal set of parameters to maximize the confinement:
\begin{align}\label{eq:minimum}
\mathbb{E}[y^2(t)]_\text{min} = \frac{8k_\text{B}T}{m\omega_{\text{d}}^2}.
\end{align}
Therefore, we see that our approximation is valid when the parameters of the system are tuned close to the optimal situation, which is done in practice when the trap is loaded with nanoparticles. 

At high pressures there is no secular oscillation: the motion is highly damped and the only observed oscillation is due to the \emph{micromotion} driving at $\omega_\text{d}$~\cite{alda2016trapping}, responsible for the peak at $\omega_{\text{d}}$ we see in the plots of Fig.~\ref{fig:3}. However, the overdamped approximation cannot be straightforwardly applied. If a particle is optically trapped in a regime where viscous forces dominate over inertia (i.e. at low Reynolds numbers), the acceleration term $(m\ddot{y})$ can be neglected, and it is assumed that the particle achieves a terminal velocity instantaneously at every new position during its motion. This approximation is valid for most experiments where micro and nanoparticles are trapped either in liquid or air~\cite{jones2015optical}. However, in Paul traps inertia has a key role on the ponderomotive force, which in turn is the responsible for the trapping mechanism~\cite{landau1959classical}. Any attempt to neglect inertia would result in a vanishing ponderomotive force, and the inclusion of the noise term in the equation of motion would produce a solution where the particle diffuses away from the trap center (see \hyperref[sec:overdamped]{Appendices}). This result adds up to a number of recent works that question the validity of the overdamped approximation in the field of stochastic thermodynamics~\citep{celani2012anomalous, martinez2015adiabatic, bo2017multiple, pan2018quantifying}. Instead, the solution we present here, based on eq.~\eqref{eq:psd}, represents a situation similar to the overdamped optical trap, whose spectral density of the position is characterized by a cutoff frequency $\omega_k=k/\gamma$, where $k$ is the stiffness of the trap~\cite{roldan2014measuring}. In our description, the cutoff frequency is given by the parameter $\abs{\lambda}$, so that the effective stiffness of the damped Paul trap is given by:
\begin{equation}\label{eq:stiffness}
	k=\frac{m\epsilon^2}{2\gamma^2}=\frac{Q^2V^2}{2md^4\Gamma^2}
\end{equation}
where the first equality clarifies the key effect of the mass on the trapping mechanism, i.e., a massless particle experiences zero restoring or ponderomotive force.

Fig.~\ref{fig:3} (a) 
shows a measurement of the motion PSD of the process (in blue) in the radial 
direction, and Fig.~\ref{fig:3}~(b) the PSD of a simulated trace (also in blue) in a low frequency regime which is hard to detect experimentally. Both plot types are compared with the OU model (eq.~\eqref{eq:psd}, red-dashed line) on a log-log scale. The measurements of the trapped particle show a corner frequency around 400 Hz, after which the spectrum decays as $1/\omega^2$. The red line in Fig.~\ref{fig:3}~(a) is a fit to eq.~\eqref{eq:psd}, which can be used to calibrate the particle's motion at ambient pressure. Experimentally, careful consideration is required on how this measurement is performed. A common motion detection technique, also adopted in this work, consists in focusing a laser on the particle and collecting the scattering light. This induces an optical force which can noticeably modify the dynamics. In our experiment, we have resorted to an extremely low effective NA to minimize the effect of the dipole potential. The simulation in Fig.~\ref{fig:3}~(c) of the process is compared to the model without any free parameters. The agreement is excellent except for the micromotion peak at $\omega_\text{d}$, which is averaged out in the model (as is expected when using an OU model). The peak's energy is however a negligible contribution to the total $\mathbb{E}[y^2(t)]$ and, hence, the PSD expression can be safely used for calibration.

\begin{figure}
\begin{center}
\includegraphics[width=0.42\textwidth]{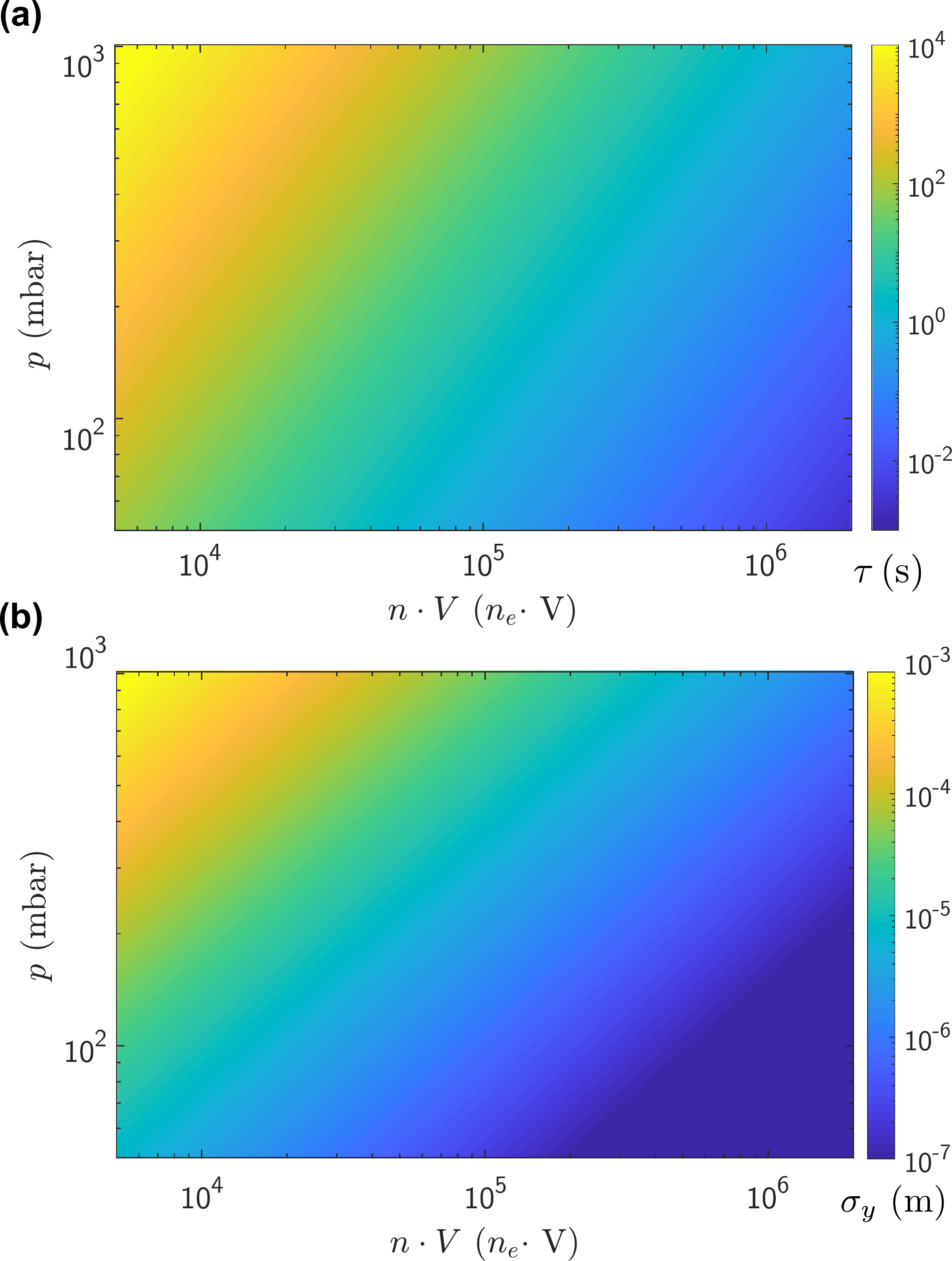}
\caption{\textbf{$\sqrt{\mathbb{E}[y^2(t)]}$ in equilibrium} (a) Dependence of the characteristic time to reach equilibrium, $\tau = 1/|\lambda|$, with the pressure $p$ and trap amplitude $Q\cdot V$, for a particle with diameter $d = 200$ nm (initial conditions $(x,v) = (0,0)$). In the plot axis, $n_e$ is the trapped particle's net number of electron charges. (b) Dependence of the confinement $\sigma_y \triangleq \sqrt{\mathbb{E}[y^2(t)]}$ with the pressure $p$ and trap amplitude $Q\cdot V$ after equilibrium has already been reached (i.e., $t \gg \tau$), for a particle with diameter $d = 200$ nm. The confinement scales with the mass as $1/\sqrt{m}$, as can be seen from eq.~\eqref{eq:equilibrium}.}\label{fig:4}
\end{center}
\end{figure}

Fig.~\ref{fig:4} displays the dependence of $\sigma_y$ and the thermalization characteristic time with the parameters that are accessible in the laboratory, namely pressure and voltage. The functional dependence was introduced in equations~\eqref{eq:lambda} and~\eqref{eq:equilibrium}, and, as we already mentioned, a relevant absence in those expressions is the trap frequency $\omega_\text{d}$, which has a negligible effect in the particle confinement between 50 mbar and ambient pressure. At the same time, $\sigma_y$ is approximately proportional to the pressure.

\section{Conclusions}
In conclusion, we have solved the general stochastic differential equation of a nanoparticle in a Paul trap, found a calibration method for the motion PSD of a levitated nanoparticle at ambient/low vacuum pressures, and applied it to a real experiment. The PSD expression has been obtained by obtaining an alternative (approximate) solution to the equation of motion with the WKB method. With it, we have proved that $y(t)$ resembles an OU process at pressures close to ambient, and studied $\langle y^2(t) \rangle$ for short and long times.  

We have complemented our theory results with thorough numerical experiments that validate the derived analytical expressions. We have also used our code to study the dependence of the particle confinement with the Paul trap parameters. 

Our findings caution against naïvely assuming that the effective potential model can be applied to any set of experimental parameters. Also, as we have found out, particles may equilibrate to the thermal bath very slowly; assuming that they thermalize faster than measurement time without the proper checks might lead to wrong conclusions. 

If these two points are under control, the next step for any levitation experiment is the calibration of the particle's motion, and the results of this work provide a novel method to do so for particles in Paul traps. Our findings can also be readily applied to estimate and optimize the particle confinement through expression~\eqref{eq:equilibrium}, which would be especially relevant in experiments that crucially depend on the particles' signal collection.

\vspace{0.5cm}

\vspace{0.5cm}
\textbf{Acknowledgments}.\hspace{0.2cm}G.P.C. and R.A.R. acknowledge financial support from the European Research Council through grant QnanoMECA (CoG - 64790), Fundació Privada Cellex, CERCA Programme / Generalitat de Catalunya, Junta de Andalucía for the project P18-FR-3583 and the Spanish Ministry of Economy and Competitiveness through the Juan de la Cierva-Incorporación grant IJCI-2015-26091, the grant PGC2018-098770-B-I00, and the Severo Ochoa Programme for Centres of Excellence in R$\&$D (SEV-2015-0522), grant FIS2016-80293-R. D.N. and J.W. were partially supported by the NSF grant DMS 1615045.

\bibliographystyle{apsrev4-1}
\bibliography{./references}

\cleardoublepage
\section{Appendices}
\phantomsection
\subsection{Explicit solution of a linear SDE}\label{sec:explicit_solution}
A linear SDE takes the general form
\begin{align}
\dif X_t = \left(A(t)X_t + a(t)\right)\dif t + \sigma(t)\dif W_t.
\end{align}
In the scalar case, $A(t)$ and $a(t)$ are numerical coefficients; in the general multivariate case $A(t)$ is a matrix and $a(t)$ is a vector. In the scalar case, we may find the solution by the following transformation and Itô's lemma:
\begin{small}
\begin{align*}
\dif \left(e^{-\int A(t)\dif t}X_t \right) & = -A(t)e^{-\int A(t)\dif t}X_t\dif t + e^{-\int A(t)\dif t}\dif X_t \\
& = e^{-\int A(t)\dif t}\left(-A(t)X_t\dif t + \dif X_t\right).
\end{align*}
\end{small}
Therefore
\begin{align*}
e^{\int A(t)\dif t}\cdot \dif \left(e^{-\int A(t)\dif s}X_t \right) = \\ \dif X_t - A(t)X_t\dif t = a(t)\dif t + \sigma(t)\dif W_t  
\end{align*}
and
\begin{align*}
\dif \left(e^{-\int A(t)\dif t}X_t \right) = e^{-\int A(t)\dif t}\left(a(t)\dif t + \sigma(t)\dif W_t \right),
\end{align*}
so an explicit solution to the SDE will be
\begin{small}
\begin{align}\label{eq:explicit}
\begin{split}
& X_t  = e^{\int_0^t A(t)\dif t}\cdot X_0 + e^{\int_0^t A(t)\dif t}\\
& \cdot\left( \int_0^t e^{-\int A(s)\dif s}a(s)\dif s + \int_0^t e^{-\int A(s)\dif s}\sigma(s)\dif W_s \right)
\end{split}
\end{align}
\end{small}

\phantomsection
\subsection{Mathieu functions}\label{sec:mathieu}
The Mathieu functions are solutions of Mathieu's differential equation
\begin{align}\label{eq:mathieu}
\ddot{y} + (a - 2q\cos(2\tau))y = 0,
\end{align}
where $a$ and $q$ are constant parameters. Stability of the solutions depends on the values of these parameters. Since it is a linear 2nd order differential equation, two independent solutions generate the linear space of all solutions of the homogeneous problem. 

In this paper, the Mathieu functions $\cm_{a,q}(\tau)$ and $\sm_{a,q}(\tau)$ are defined as the solutions of~\eqref{eq:mathieu} with initial conditions $(y, \dot{y}) = (1, 0)$ (for $\cm_{a,q}(\tau)$) and $(y, \dot{y}) = (0, 1)$ (for  $\sm_{a,q}(\tau)$), in clear analogy to the standard $\cos(\tau)$ and $\sin(\tau)$ functions. 

For a detailed discussion of the Mathieu equation and functions' properties, see \url{http://dlmf.nist.gov/28.2}.

\phantomsection
\subsection{Experimental parameters}\label{sec:parameters}
We use the following set of parameters, based on our own experimental setup:
\begin{itemize}
\item $T = 295$ K (ambient temperature).
\item Particle radius: we assume $100 \leq r \leq 1000$ nm. From the radius, $m = \frac{4}{3}r^3\cdot \rho$, where $\rho$ is the material density. Assuming silica, this results in $m  \in [10^{-17} \, 10^{-15}]$ kg.
\item $\gamma = 6\pi \nu r$, where $\nu = 18.6\cdot 10^{-6}$ Pa$\cdot$s is the viscosity of air at ambient pressure. This gives $\gamma = 3.5 \cdot 10^{-11}$ kg/s. $\sigma = \sqrt{2k_\text{B} T \gamma}$ is obtained from the fluctuation-dissipation theorem, $\sigma = 5.3 \cdot 10^{-16}$.
\item $\omega_\text{d}/2\pi \in [0.5,\, 20]$ kHz
\item $Q$ is the net number of charges in the particle (assume $50 \leq Q \leq 1000$), $V$ the electric potential at the electrodes ($500 \leq V \leq 2000$ volts) and $d^2$ a constant factor that takes into account the geometry of the trap ($0.1 < d < 1$ mm). From these we calculate $\epsilon = \frac{Q V}{d^2}$. $\epsilon \in [4\cdot 10^{-9} \quad 3\cdot 10^{-5}]$
\end{itemize}

\phantomsection
\subsection{Naïve overdamped approximation}\label{sec:overdamped}
If inertia $(m\ddot{y})$ is neglected, Equation~\ref{eq:motion} becomes
\begin{align}\label{eq:motion_over}
\gamma \dot{y} - \epsilon \cos \omega_\text{d} t\cdot y = \sigma \eta(t), 
\end{align}
which in the main text is referred as the \emph{naïve} overdamped approximation. The solution to this equation is: 
\begin{align}\label{eq:sol_over}
y(t)=e^{-\frac{\epsilon}{\gamma\omega_\text{d}}\sin\omega_\text{d}t}\left(y(0)+\frac{\sigma}{\gamma}\int_0^te^{\frac{\epsilon}{\gamma\omega_\text{d}}\sin\omega_\text{d}r}\text{d}W_r\right)
\end{align}
Assuming $y(0) = 0$ for simplicity, we may apply Itô's isometry in \eqref{eq:sol_over}, finding the following expression for the process variance:
\begin{align}\label{eq:variance_over}
\mathbb{E}[y^2(t)] =e^{-\frac{2\epsilon}{\gamma\omega_\text{d}}\sin\omega_\text{d}t}\left(\frac{\sigma^2}{\gamma^2}\int_0^te^{\frac{2\epsilon}{\gamma\omega_\text{d}}\sin\omega_\text{d}s}\text{d}s\right)
\end{align}
which grows with time as $\mathcal{O}(t)$. This behaviour is qualitatively different from the one obtained with the solution to the full equation and indicates that, in this case, the singular perturbation problem requires a more sophisticated approach than just neglecting the highest order term. 

Some physical intuition on why the approximation fails can be gained noticing that in the absence of inertia or noise, no net force acts on average over a cycle of the driving field. By linearity, superimposing a Brownian motion leads to a uniformly growing variance, thus avoiding a stable trapping regime.

\phantomsection
\subsection{WKB calculations}\label{sec:wkb}

We study the dynamics of a trapped particle, satisfying the equation of motion
\begin{align}\label{eq:supp_1}
m\ddot{y} + \gamma\dot{y} - \epsilon \cos(\omega_\text{d} t) y = \sigma \eta_t
\end{align}
Here, $\gamma$, $\epsilon$, $\omega_\text{d}$ and $\sigma$ are constants and $\eta_t$ denotes a unit intensity white noise; $\gamma$ is related to the $\Gamma$ of the main text by $\gamma = \Gamma m$. $m$ is the particle's mass and we are interested in small values of $m$.  We are thus going to conduct an asymptotic analysis of the solutions to \eqref{eq:supp_1} as $m \to 0$.  The primary object of study is the variance of the particle's position, i.e. the expected value of $\left(y(t) - \mu_t\right)^2$ where $\mu_t$ denotes the expected value of $y(t)$.  In the case when $y(0) = \dot{y}(0) = 0$, we have $\mu_t = 0$ and the variance is equal to the second moment $\mathbb{E}\left[y(t)^2\right]$.  Here (and in all the text) $\mathbb{E}$ denotes expected value.

First, we are going to study the homogeneous equation:
\begin{align}\label{eq:supp_2}
m\ddot{y} + \gamma\dot{y} - \epsilon \cos(\omega_\text{d} t) y = 0.
\end{align}
This second-order equation has two linearly independent solutions.  We want to find their approximate expressions, in order to study the solution of the randomly perturbed equation \eqref{eq:supp_1}. To do this we first postulate that $y(t)$ has the form
\begin{align}\label{eq:supp_3}
y(t) = \exp\left(-\frac{\gamma }{ 2m}t\right)u(t)
\end{align}
and substitute into \eqref{eq:supp_2} obtaining
\begin{align}\label{eq:supp_4}
m\ddot{u} = \left(\frac{\gamma^2 }{ 4m} + \epsilon \cos(\omega_\text{d} t)\right)u.
\end{align}
Multiplying both sides by $m$, we rewrite \eqref{eq:supp_4} as
\begin{align}\label{eq:supp_5}
m^2\ddot{u} = \left(\frac{\gamma^2 }{ 4} + m\epsilon \cos(\omega_\text{d} t)\right)u
\end{align}
We will study the last equation, by the WKB method. Using the ansatz
\begin{align}\label{eq:supp_6}
u(t) = \exp\left(\frac{1 }{ m}S_0(t) + S_1(t) + mS_2(t)+\dots\right)   
\end{align}
we get
\begin{align}\label{eq:supp_7}
\begin{split}
\ddot{u} & = \left(\frac{1 }{ m}\ddot{S}_0 +  \ddot{S}_1 + m\ddot{S}_2 + \dots\right)u \\
& + \left(\frac{1 }{ m}\dot{S}_0 + \dot{S}_1 + m\dot{S}_2 + \dots\right)^2u.
\end{split}
\end{align}
Substituting this formula into equation \eqref{eq:supp_5} and dividing by $u$ we obtain
\begin{align}\label{eq:supp_8}
\begin{split}
m^2 \left(\frac{1 }{ m}\ddot{S}_0 + \ddot{S}_1 + m\ddot{S}_2 + \dots\right) +  \\ 
m^2\left(\frac{1 }{ m}\dot{S}_0 + \dot{S}_1 + m\dot{S}_2 + \dots\right)^2 = \frac{\gamma^2}{4} + m\epsilon \cos(\omega_\text{d} t)
\end{split}
\end{align}
We now compare the coefficients of the same powers of $m$ on both sides of this equation.
In order $m^0$ we obtain
\begin{align}\label{eq:supp_9}
\left(\dot{S_0}\right)^2 = \frac{\gamma^2 }{ 4}
\end{align}
This equation has two solutions, $\dot{S_0} = \frac{\gamma }{ 2}$ and $\dot{S_0} = -\frac{\gamma }{ 2}$.  Let us start from the first case, so that 
\begin{align}\label{eq:supp_10}
S_0(t) = \frac{\gamma }{ 2}t
\end{align}
Comparing the terms proportional to $m^1$ on both sides of \eqref{eq:supp_8}, we obtain
\begin{align}\label{eq:supp_11}
\ddot{S_0} + 2\dot{S_0}\dot{S_1} = \epsilon\cos(\omega_\text{d} t)
\end{align}
Substituting $S_0(t)$ from \eqref{eq:supp_10} we obtain
\begin{align}\label{eq:supp_12}
S_1(t) = \frac{\epsilon }{ \gamma \omega_\text{d}}\sin(\omega_\text{d} t)
\end{align}
Next, comparing the terms proportional to $m^2$ on both sides of \eqref{eq:supp_8}, we find
\begin{align}\label{eq:supp_13}
\ddot{S_1} + \left(\dot{S_1}\right)^2 + 2\dot{S_0}\dot{S_2} = 0
\end{align}
Substituting the calculated expressions for $S_0$ and $S_1$ from \eqref{eq:supp_10} and \eqref{eq:supp_12} and solving for $\dot{S_2}$ we obtain:
\begin{align}\label{eq:supp_14}
\dot{S}_2 = \frac{\epsilon \omega_\text{d}}{\gamma^2}\sin(\omega_\text{d} t) - \frac{\epsilon^2}{2\gamma^3}\left(1 + \cos(2\omega_\text{d} t)\right)
\end{align}
Integrating, we get 
\begin{align}\label{eq:supp_15}
S_2(t) = -\frac{\epsilon}{\gamma^2}\cos(\omega_\text{d} t) - \frac{\epsilon^2 }{ 2\gamma^3}t - \frac{\epsilon^2 }{ 4\gamma^3 \omega_\text{d}}\sin(2\omega_\text{d} t)
\end{align}
Substituting the derived expressions for $S_0, S_1$ and $S_2$ into \eqref{eq:supp_6} and neglecting the higher order terms of the series in the exponent, we obtain for the first of two linearly independent solutions of \eqref{eq:supp_5} the approximate formula
\begin{align}\label{eq:supp_16}
\begin{split}
u(t) \approx \exp\left[\left(\frac{\gamma }{ 2m} - \frac{m\epsilon^2 }{ 2\gamma^3}\right)t + \frac{\epsilon }{ \gamma \omega_\text{d}}\sin(\omega_\text{d} t) - \right. \\
\left. \frac{m\epsilon }{ \gamma^2} \cos(\omega_\text{d} t) -\frac{m\epsilon^2 }{ 4\gamma^3\omega_\text{d}}\sin(2\omega_\text{d} t)\right].
\end{split}
\end{align}
We now multiply $u(t)$ by $\exp\left(-\frac{\gamma }{ 2m}t\right)$ and drop from the expression in the exponent of \eqref{eq:supp_16} the periodic terms proportional to $m$ (a fuller justification of this step will be given below) to obtain an approximate expression for the first of two linearly independent solutions of the equation \eqref{eq:supp_2}:
\begin{align}\label{eq:supp_17}
y_1(t) \approx \exp\left[\left( - \frac{m\epsilon^2}{2\gamma^3}\right)t + \frac{\epsilon}{ \gamma \omega_\text{d}}\sin(\omega_\text{d} t) \right].
\end{align}
An approximation of the second one follows from the choice 
\begin{align}\label{eq:supp_18}
S_0 = -\frac{\gamma}{2}t
\end{align}
instead of \eqref{eq:supp_10}. A calculation fully analogous to the one presented above leads to
\begin{align}\label{eq:supp_19}
y_2(t) \approx \exp\left[\left(-\frac{\gamma }{ m} + \frac{m\epsilon^2}{2\gamma^3}\right)t  - \frac{\epsilon}{\gamma \omega_\text{d}}\sin(\omega_\text{d} t)\right].
\end{align}
Our approximate formulae are in agreement with Floquet theory, according to which two linearly independent solutions of eqn. Eq. \eqref{eq:supp_2} can be chosen in the form
\begin{align}\label{eq:supp_20}
y_j(t) = \exp\left(\lambda_j t\right)P_j(t)
\end{align}
for $j = 1,2$, where $P_j$ are $\frac{2\pi }{ \omega_\text{d}}$-periodic functions.  $\lambda _1$ and $\lambda_2$ are the \emph{characteristic exponents} of the equation.  The above statement is true whenever the characteristic exponents are distinct, which is clearly the case here.  In what follows, only the values of $\lambda_j$ are going to play a role, since the periodic factors $P_j$ are bounded.  This is the reason why we could omit corrections of order $m$ to the exponents in \eqref{eq:supp_17} and \eqref{eq:supp_17}. Explicitly, we have seen that, according to the WKB approximation:
\begin{align}\label{eq:supp_21}
\lambda_1 &\approx - \frac{m\epsilon^2 }{ 2\gamma^3} \\  
\lambda_2 &\approx -\frac{\gamma }{ m} + \frac{m\epsilon^2 }{ 2\gamma^3} \\
P_1(t) &= \exp\left(\frac{\epsilon }{ \gamma\omega_\text{d}}\sin(\omega_\text{d} t)\right) \\
P_2(t) &= \exp\left(-\frac{\epsilon }{ \gamma\omega_\text{d}}\sin(\omega_\text{d} t)\right)
\end{align}
We now use these values to study the behavior of solutions to the randomly perturbed equation \eqref{eq:supp_1}. This can be done directly, using variation of constants method for the inhomogeneous second order equation \eqref{eq:supp_1}. According to this method, \eqref{eq:supp_1} has a particular solution equal to
\begin{align}
y(t) = \int_0^t\Delta(s)^{-1}\left[y_1(s)y_2(t) - y_1(t)y_2(s)\right]\frac{\sigma }{ m}\eta_s\,\dif s
\end{align}
where 
\begin{align}
\Delta(s) = y_1(s)\dot{y}_2(s) - \dot{y}_1(s)y_2(t).
\end{align}
Below we derive the same expression in a different way. First, we rewrite our equation as a system:
\begin{align}\label{eq:supp_22}
\dot{y} &= v \\
\dot{v} &= \frac{\epsilon }{ m}y -\frac{\gamma }{ m}v + \frac{\sigma }{ m}\eta_t 
\end{align}
Introducing the vector of dynamical variables
\begin{align}\label{eq:supp_23}
\textbf{x} = \begin{pmatrix}
y \\
v
\end{pmatrix}
\end{align}
the matrix
\begin{align}\label{eq:supp_24}
\textbf{A}(t) = \begin{pmatrix}
0 &1 \\ 
\frac{\epsilon}{m}\cos(\omega_\text{d} t) &-\frac{\gamma }{ m} 
\end{pmatrix}
\end{align}
and the noise vector
\begin{align}\label{eq:supp_25}
\textbf{h}(t) = \begin{pmatrix}
0 \\
\frac{\sigma}{m}\eta_t
\end{pmatrix}
\end{align}
we can rewrite the above system as
\begin{align}\label{eq:supp_26}
\dot{\textbf{x}}(t) = {\textbf{A}}(t){\textbf x}(t) + {\textbf h}(t)
\end{align}
If $y_1$ and $y_2$ are two linearly independent solutions of the homogeneous equation \eqref{eq:supp_2}, then the matrix
\begin{align}\label{eq:supp_27}
\textbf{Q}(t) = \begin{pmatrix}
y_1(t) &y_2(t) \\
\dot{y}_1(t) & \dot{y}_2(t)
\end{pmatrix}
\end{align}
is a fundamental matrix of the homogeneous system
\begin{align}\label{eq:supp_28}
\dot{\textbf{x}}(t) = \textbf{A}(t)\textbf{x}(t)
\end{align}
i.e. $\textbf{Q}(t)$ satisfies the matrix ODE
\begin{align}\label{eq:supp_29}
\dot{\textbf{Q}}(t) = \textbf{A}(t)\textbf{Q}(t)
\end{align}
The solution of the inhomogeneous system can be written as
\begin{align}\label{eq:supp_30}
\textbf{x}(t) = \textbf{Q}(t)\textbf{Q}(0)^{-1}\textbf{x}(0) + \int_0^t \textbf{Q}(t)\textbf{Q}(s)^{-1}\textbf{h}(s)\,\dif s
\end{align}
This follows from the variation of constants formula and can be easily verified by direct differentiation.
Note that because both characteristic exponents are negative, the entries of $\textbf{Q}(t)$ decay exponentially, and so does the first term in the above formula.  Note also, that if the particle is initially at $y = 0$ with zero velocity, than $\textbf{x}(0) = 0$ and this first term vanishes. In any case, the asymptotic behaviour of the solution is determined by the second term.
\begin{widetext}
We have 
\begin{align}\label{eq:supp_31}
\textbf{Q}(t) = \begin{pmatrix}
P_1(t)e^{\lambda_1t} & P_2(t)e^{\lambda_2t} \\
\dot{P}_1(t)e^{\lambda_1 t} + \lambda_1 P_1(t)e^{\lambda_1t} &\dot{P}_2(t)e^{\lambda_2t} + \lambda_2P_2(t)e^{\lambda_2t}
\end{pmatrix}
\end{align}
Hence 
\begin{align}\label{eq:supp_32}
\Delta(t) := \det {\textbf Q}(t) = -e^{(\lambda_1 + \lambda_2)t}\left[\left(\lambda_1 - \lambda_2\right)P_1(t)P_2(t) - \left(P_1(t)\dot{P}_2(t) - \dot{P}_1(t)P_2(t)\right)\right]   
\end{align}
To leading order we thus have
\begin{align}\label{eq:supp_33}
\Delta(t)  \approx -(\lambda_1 - \lambda_2)e^{(\lambda_1 + \lambda_2)t}P_1(t)P_2(t) \sim -\frac{\gamma }{ m}\exp\left(-\frac{\gamma }{ m}t\right)
\end{align}
Now, 
\begin{align}\label{eq:supp_34}
\textbf{Q}(s)^{-1} = \Delta(s)^{-1}
\begin{pmatrix}
\dot{P}_2(s)e^{\lambda_2s} + \lambda_2P_2(s) e^{\lambda_2s} &-P_2(s)e^{\lambda_2s} \\
-\dot{P}_1(s)e^{\lambda_1s} - \lambda_1P_1(s)e^{\lambda_1s} &P_1(s)e^{\lambda_1s}
\end{pmatrix}
\end{align}
We want to study the first component of the vector $\textbf{x}(t)$ in $(30)$
Since the first component of the noise vector $\textbf{h}(s)$ in $(25)$ equals $0$, to calculate the first component of the integral in \eqref{eq:supp_30}, we need to multiply the $(1,2)$-element of the matrix $\textbf{Q}(t)\textbf{Q}(s)^{-1}$ by $\frac{\sigma}{m}\eta_s$ and integrate over $s$ from $0$ to $t$.  The $(1,2)$-matrix element of $\textbf{Q}(t)\textbf{Q}(s)^{-1}$ equals
\begin{align}\label{eq:supp_35}
\left(\textbf{Q}(t)\textbf{Q}(s)^{-1}\right)_{1,2} = \Delta(s)^{-1}\left[-P_1(t)P_2(s)e^{\lambda_1t + \lambda_2s} + P_2(t)P_1(s)e^{\lambda_2t + \lambda_1s}\right]
\end{align}
\end{widetext}
Approximating $\Delta(s)$ by its leading term, according to \eqref{eq:supp_33}, we obtain
\begin{align}\label{eq:supp_36}
\left({\textbf Q}(t){\textbf Q}(s)^{-1}\right)_{1,2} \approx \\
\frac{m}{\gamma}\left[ \frac{P_1(t) }{ P_1(s)}e^{\lambda_1(t-s)} - \frac{P_2(t)}{P_2(s)}e^{\lambda_2(t-s)}\right]
\end{align}
The factors ${P_1(t) / P_1(s)}$ and ${P_2(t) / P_2(s)}$ are of order $1$ and, in fact, quite close to $1$, since for the considered values of the parameters, $\frac{\epsilon }{ \gamma \omega_\text{d}}$ is of the order of $10^{-3}$ (see \eqref{eq:supp_21}).  Of the two exponential factors, $e^{\lambda_1(t-s)}$ and $e^{\lambda_2(t-s)}$, the second decays much faster with $t-s$ and is negligible for small $m$.  We are thus left with the following approximation to $y(t)$ (for simplicity we consider the initial conditions $y(0) = v(0) = 0$):
\begin{align}\label{eq:supp_37}
y(t) \approx \frac{\sigma }{ \gamma}\int_0^te^{\lambda_1(t-s)}\eta_s\,\dif s = \frac{\sigma }{ \gamma}\int_0^te^{\lambda_1(t-s)}\,dW_s  
\end{align}
where the last expression is a stochastic integral with respect to the Wiener process $W_t$, formally (and in the sense of distribution theory) satisfying $\frac{\dif W_t }{\dif t} = \eta_t$.  The It\^o isometry implies that 
\begin{align}\label{eq:supp_38}
\mathbb{E}\left[y(t)^2\right] \approx \left(\frac{\sigma }{ \gamma}\right)^2\int_0^te^{2\lambda_1(t-s)}\,\dif s 
\end{align}
Extending the integral on the right-hand side of the above equation to infinity, we obtain an approximate bound
\begin{align}\label{eq:supp_39}
\mathbb{E}\left[y(t)^2\right] \lesssim \frac{\sigma^2 }{ 2\gamma^2 |\lambda_1|} = \frac{\sigma^2\gamma }{ m\epsilon^2}
\end{align}
Rather than substituting specific values of the physical parameters, we compare (the square root of) this result to the experimental value of the standard deviation in order to estimate the value of the particle's actual mass.

We note that, since the exponent $\lambda_1$ is very close to zero, for small times $t$ the integrand $e^{2\lambda_1 t}$ is close to $1$ and the approximate value of the variance is
\begin{align}\label{eq:supp_40}
\mathbb{E}\left[y(t)^2\right] \approx \frac{\sigma^2 }{ \gamma^2}t 
\end{align}
This is a good approximation (first term of the Taylor expansion) as long as $\lambda_1 t \ll 1$, thus introducing a time scale $\tau = \frac{1 }{ \lambda_1}$.  For times much smaller than $\tau$ the variance grows approximately linearly with $t$, so the equation describes diffusion; for times larger than $\tau$, localization effects become pronounced.

It is interesting to compare the results of the above calculation with those obtained from a different limit, used in the reference~\citep{izmailov1995microparticle}.

In the second case, with the new small parameter, we start by dividing eq. \eqref{eq:newequation} by $\kappa^2$ and transforming it into the system
\begin{align}
\vec{\dot{x}}(\tau)  =  \mathbf{A}(\tau) \vec{x}+ \vec{f}(\tau),
\end{align}
where
\begin{align*}
\vec{x} = \left( {\begin{array}{cc}
	x(t)  \\
	\dot{x}(t)
	\end{array}} \right)&,  \quad \mathbf{A} =
\left( {\begin{array}{cc}
	0 & 1 \\
	\beta' \cos(2\tau)/\kappa^2 & - 1 / \kappa  
	\end{array}} \right),  \\
&\vec{f}(\tau) = \left( {\begin{array}{cc}
	0  \\
	\frac{4 \sigma}{m \omega_\text{d}^2} \sqrt{\frac{\omega_\text{d}}{2}}  \eta(\tau) 
	\end{array}} \right).
\end{align*}

Let $\mathbf{\Phi}(\tau) $ be a principal fundamental matrix solution when $ \vec{f}(\tau)=\vec{0}$. Then the general solution is
\begin{align}
\vec{x} (\tau)=   \mathbf{\Phi}(\tau) \vec{x}(0) + \int_{0}^{\tau} \mathbf{\Phi}(\tau) {\mathbf{\Phi}(s)}^{-1}  \vec{f}(s)\,\dif s.
\end{align}

From the previous derivation, the two linearly independent solutions of \eqref{eq:newequation} are
\begin{equation}
\begin{split}
x_{1,2}&(\tau) = \left(\frac{1}{4} + \beta' \cos(2\tau)\right)^{-\frac{1}{4}} \\
& \cdot\exp \left( -\frac{\tau}{2 \kappa} \pm \frac{1}{\kappa} \int_{0}^{\tau} \sqrt{\frac{1}{4} + \beta' \cos(2\tau)}\,\dif \tau \right).
\end{split}
\end{equation}

Taylor expanding this expression~\footnote{This is justified because $\beta' \ll 1$} gives
\begin{small}
\begin{align}
\begin{split}
&x_{1,2}(\tau)=\left(\frac{1}{4}\right)^{-\frac{1}{4}}\cdot \exp \left(- \frac{\tau}{2 \kappa}\right) \\
& \cdot \exp\left(\pm \frac{1}{\kappa} \int_{0}^{\tau}\left( \frac{1}{2} + \beta' \cos(2\tau) - \beta'^2 \frac{1+\cos(4\tau)}{2} \,\dif \tau \right) \right).
\end{split}
\end{align}
\end{small}

This results in
\begin{small}
\begin{equation}\label{eq:afterintegral}
\begin{split}
&x_{1,2}(\tau)=\left(\frac{1}{4}\right)^{-\frac{1}{4}}\cdot \exp \left(- \frac{\tau}{2 \kappa}\right) \\
&\cdot \exp\left( \pm \left( \frac{\tau}{2 \kappa} + \frac{\beta' \sin(2\tau)}{2 \kappa} - \frac{{\beta'}^2 \tau}{2\kappa} -\frac{(\beta')^2 \sin(4 \tau )}{8 \kappa} \right) \right).
\end{split}
\end{equation}
\end{small}
Using  $\tau= \frac{\omega_\text{d} t}{2}, \beta'=\frac{m \epsilon}{\gamma^2}, \kappa= \frac{m \omega_\text{d}}{2 \gamma}$, and substituting into eq. \eqref{eq:afterintegral}, we find (approximate) explicit expressions for the two fundamental solutions,
\begin{equation}
x_1 = \left(\frac{1}{4}\right)^{-\frac{1}{4}}\cdot \exp \left(  \frac{\epsilon \sin(\omega_\text{d} t)}{\omega_\text{d} \gamma} -\frac{  m \epsilon^2 t}{ \gamma^3}  \right)
\end{equation}
and
\begin{equation}
x_2 =\left(\frac{1}{4}\right)^{-\frac{1}{4}}\cdot \exp \left( - \frac{ \gamma t}{m} -  \frac{\epsilon \sin(\omega_\text{d} t)}{\omega_\text{d} \gamma}\right).
\end{equation}
If $P_1(\tau)e^{\lambda_1 \tau}$, $P_2(\tau)e^{\lambda_2 \tau}$ represent two linearly independent solutions of the homogeneous equation, then
\begin{widetext}
\begin{align}
\left( \mathbf{\Phi}(\tau) \mathbf{\Phi}^{-1}(s) \right)_{1,2}& = {\frac{e^{\lambda_2 (\tau-s)}P_2(\tau)P_1(s) -e^{\lambda_1 (\tau-s)}P_1(\tau)P_2(s)}  {   P_1(s) (P'_2(s) + \lambda_2 P_2(s))- P_2(s)(P'_1(s) + \lambda_1 P_1(s))  }} \\
& \approx \frac{m \omega_\text{d}}{2 \gamma} e^{\frac{-  m \epsilon^2 (\tau-s) }{\omega_\text{d} \gamma^3}}.
\end{align}
\end{widetext}

We still need to solve the inhomogeneous equation, for which we compute 
\begin{align}\label{eq:explicit_integral}
\int_{0}^{\tau} & \left(\mathbf{\Phi}(\tau) \eta(s) {\mathbf{\Phi}(s)}^{-1}\right)_{1,2}  f_2(s) \eta(s) \,\dif s \nonumber \\
& \approx \int_{0}^{\tau} \frac{m \omega_\text{d}}{2 \gamma}       \frac{4 \sigma}{m \omega_\text{d}^2} \sqrt{\frac{\omega_\text{d}}{2}}  \eta(s) e^{\frac{-  m \epsilon^2 (\tau-s) }{\omega_\text{d} \gamma^3}} \eta(s) \,\dif s \nonumber\\
& = \int_{0}^{\tau} \frac{\sigma}{\gamma}  \sqrt{\frac{2}{\omega_\text{d}}} e^{\frac{-  m \epsilon^2 (\tau-s) }{\omega_\text{d} \gamma^3}} \eta(s) \,\dif s.
\end{align}

This provides us with an explicit solution of the process $x(t)$ in terms of elementary functions, and we may use it to calculate the variance of the process~\footnote{The first moment, $\mathbb{E}[x(t)] = 0$ by the properties of the Itô integral with respect to a Wiener process.}
Applying  the Itô isometry to eq. \eqref{eq:explicit_integral}, we get
\begin{align} 
\mathbb{E}\left[ \int_{0}^{\tau} \frac{\sigma}{\gamma} \sqrt{\frac{2}{\omega_\text{d}}} e^{\frac{-  m \epsilon^2 (\tau- s) }{\omega_\text{d} \gamma^3}} \dif W_s \right]^2 \\
= \int_{0}^{\tau} \frac{\sigma^2}{\gamma^2} \frac{2}{\omega_\text{d}} e^{\frac{- 2 m \epsilon^2 (\tau-s) }{\omega_\text{d} \gamma^3}} \dif s 
\end{align}

Using $s= \frac{\omega_\text{d} t}{2}$, we finally obtain 
\begin{align}
\mathbb{E}[y^2(t)] \simeq \left(\frac{\sigma}{\gamma}\right)^2 \int_{0}^{\frac{ 2 t}{\omega_\text{d} }}    e^{\frac{-m \epsilon^2 t}{\gamma^3}} \dif s.
\end{align}
This is in agreement with what was obtained from the small mass limit.

\phantomsection
\subsection{Approximate spectral density}\label{sec:psd}
We now use the approximate solution of the equation of motion to calculate the (also approximate) spectral density.  That is, we put
\begin{align}
x_t = \frac{\sigma }{ \gamma}\int_0^te^{\lambda(t-s)}\,\dif W_s.
\end{align}
For simplicity, we assumed that $x(0) = \dot{x}(0) = 0$; the following calculation can be easily generalized to the case of arbitrary initial conditions. We use the definition of the spectral density:
\begin{align}\label{eq:psd_expression}
S(\omega) = \lim_{T \to \infty}\frac{1 }{ T}\,\mathbb{E}\left|\int_0^T x(t)e^{-i\omega t}\,\dif t\right|^2
\end{align}

Predictably, the result is the spectral density of the OU process, approximating the exact solution:
\begin{align}\label{eq:psd_approx}
S(\omega) = \frac{\sigma^2 }{ \gamma^2}\frac{1 }{ \omega^2 + \lambda^2}
\end{align}
For completeness, we include the expression for the expected value in \eqref{eq:psd_expression}, \emph{before} the limit is taken, thus providing corrections to \eqref{eq:psd_approx}, which go to zero as $T \to \infty$.   The calculation relies on the It\^o isometry and tedious but elementary integration.  Its result is:
\begin{align}
S_T(\omega) = \frac{\sigma^2 }{ \gamma^2}\frac{1 }{ \omega^2 + \lambda^2}\left(1 + \frac{A_T}{T}\right)
\end{align}
where 
\begin{align}
\begin{split}
A_T & = \frac{e^{2\lambda T} - 1 }{ 2 \lambda} - \\
& \frac{2}{\lambda^2 + \omega^2}\left[\lambda \left(e^{\lambda T} \cos \alpha T - 1\right) + \omega e^{\lambda T}\sin \omega T\right]
\end{split}
\end{align}
It is clear that the correction $\frac{A_T}{T}$ goes to zero, at the rate $\frac{1}{T}$ (let us remind that the characteristic exponent $\lambda$ is negative).

\phantomsection
\subsection{Details of the numerical simulation}\label{sec:numerical}

We have developed functions and libraries in C++ to generate sample paths of a vector process (of arbitrary dimension) defined by a stochastic differential equation
\begin{align*}
dif \textbf{X} = \textbf{a}(t, \textbf{X})\,\dif t+ \textbf{b}(t, \textbf{X})\,\dif W.
\end{align*}
The simulation of $X_t$ is performed with a modified Runge-Kutta method for stochastic differential equations~\citep{roberts2012modify} (strong order 1, deterministic order 2), detailed at the end of the section. This particular method does not require any non-zero derivatives of the diffusion term $\textbf{b}(t, \textbf{X})$. Other methods (e.g. the Milstein method) have strong order 1 but reduce to the Euler-Maruyama method (strong order 0.5) when $\textbf{b}(t, \textbf{X})$ is a constant.

Since the realizations of the process have a certain degree of randomness, each of them will be different and many traces (usually around $n = 1000$) need to be generated to estimate the process statistical moments 
\begin{align*}
\mathbb{E}[f(X_t)] \simeq \frac{1}{n}\sum_i f(X^i_t).
\end{align*}
This is usually quite intensive in terms of processing power and computer memory. For this reason, the main computation is coded in C++, while MATLAB and Python are used for post-processing. The code and libraries can be freely downloaded from \href{https://github.com/gerardpc/sde\textunderscore simulator}{https://github.com/gerardpc/sde\_simulator}. 

\subsection{Runge-Kutta method}

Let $\mathbf{X}(t) \in \mathbb{R}^n$ be the stochastic process satisfying the general It\^o stochastic differential equation (SDE):

Given a time step $\Delta t$ and the value $\textbf{X}(t_k)= \textbf{X}_k$, then $\textbf{X}(t_{k+1})$ is calculated recursively as
\begin{small}
\begin{align*}
\textbf{K}_1 = & \textbf{a}(t_k, \textbf{X}_k) \Delta t  +(\Delta W_k-S_k\sqrt{\Delta t})\cdot \textbf{b}(t_k, \textbf{X}_k),
\\
\textbf{K}_2 = & \textbf{a}(t_{k+1}, \textbf{X}_k+ \textbf{K}_1) \Delta t \,+\\
& (\Delta W_k+S_k\sqrt{\Delta t}) \cdot \textbf{b}(t_{k+1}, \textbf{X}_k+\textbf{K}_1),\\
\textbf{X}_{k+1} = &  \textbf{X}_k + \frac12(\textbf{K}_1 + \textbf{K}_2),
\end{align*}
\end{small}
where $\Delta W_k \sim \mathcal{N}(0, \Delta t)$, and $S_k = \pm 1$, having each probability 1/2. Setting $S_k = 0$ will approximate $X_t$ in the Stratonovich sense.

\end{document}